\documentclass[12pt]{article}
\usepackage{epsfig,rotating}
\usepackage{amssymb}
\begin{document}
\baselineskip 16pt plus 2pt minus 2pt
\newcommand{\beq}{\begin{equation}}
\newcommand{\eeq}{\end{equation}}
\newcommand{\beqa}{\begin{eqnarray}}
\newcommand{\eeqa}{\end{eqnarray}}
\newcommand{\dida}[1]{/ \!\!\! #1}
\renewcommand{\Im}{\mbox{\sl{Im}}}
\renewcommand{\Re}{\mbox{\sl{Re}}}
\newcommand{\PRD}[3]{{Phys.~Rev.} \textbf{D#1},({#2}) #3}
\newcommand{\PLB}[3]{{Phys.~Lett.} \textbf{B#1}, ({#2}) #3}
\newcommand{\PRL}[3]{{Phys.~Rev.~Lett.} \textbf{#1}, ({#2}) #3}
\newcommand{\NPB}[3]{{Nucl.~Phys.} \textbf{B#1}, ({#2}) #3}
\newcommand{\NPA}[3]{{Nucl.~Phys.} \textbf{A#1}, ({#2}) #3}
\def\simge{\hspace*{0.2em}\raisebox{0.5ex}{$>$}
     \hspace{-0.8em}\raisebox{-0.3em}{$\sim$}\hspace*{0.2em}}
\def\simle{\hspace*{0.2em}\raisebox{0.5ex}{$<$}
     \hspace{-0.8em}\raisebox{-0.3em}{$\sim$}\hspace*{0.2em}}

\begin{titlepage}



\vspace{1.0cm}

\begin{center}
{\large {\bf  Neutrinos in Extra Dimensions and the Anomalous 
Magnetic Moments of Leptons}}\\

\vspace{1.2cm}

G. C. McLaughlin\footnote{email:
gail@triumf.ca}
and
J. N. Ng\footnote{email: misery@triumf.ca}\\

\vspace{0.8cm}
TRIUMF, 4004 Wesbrook Mall, Vancouver, B.C., Canada V6T 2A3
\end{center}

\vspace{1cm}

\begin{abstract}
The use of extra dimensional scenarios as models for neutrino
mass affects many low energy observables.  We
consider the implications of virtual bulk neutrinos in
precision experiments of the
anomalous magnetic moments of the muon and the electron.  We 
consider neutrino models in factorizable geometry of the type $M_4 \times T$ as 
well as the sliced $AdS_5$ non-factorizable geometry. 
In both geometries we find finite contributions
to $g-2$ after summing over the KK excitations of the bulk neutrinos. 
In the case of Randall-Sundrum geometry, we find that the muon
experiment is approaching the precision necessary to probe
these models.
\end{abstract}  

\vspace{2cm}
\vfill
\end{titlepage}
                                                                                                               
The existence of extra dimensions is a common feature of many extensions of the Standard
Model (SM). In particular it is well known that string theory can be formulated 
consistently in
10 or 11 dimensions. Usually the higher dimensional geometry 
is taken to be $M_4 \times T$ where
$M_4$ is Minkowski space and $T$ denotes the compactified space of the extra dimensions. 
The extra dimensions are usually 
taken to be spatial and the geometry of $T$ is assumed to consist of 
 spheres or tori with small radii.
Therefore, the extra dimensions avoid detection in current experiments. Recently, it
has been suggested that 
extra dimensions can be used to explain the hierarchy between
the weak and the Planck scale. There are two distinct scenarios
which implement this idea. 
In one case the extra dimensions are taken to be large \cite {ADD,AADD} and the 
apparent hierarchy is due to the large volume of the extra dimensions. 
Specifically, the 
fundamental scale $M_*$ in $4+n$ dimensions
where $n$ denotes the number of extra dimensions is 
related to the 4D Planck scale $M_{Pl}$ via the relation
\beq
\label {4DPl}
  M_{Pl}^2=M_*^{n+2}V_n
\eeq
and $V_n$ is the volume of the compactified extra dimensions. For simplicity 
we shall take the 
geometry of this compactification to be a $n-$dimensional
torus with equal radii $R$ and 
thus $V_n = (2\pi R)^n$. If
$R$ is of the order of 1 mm then $M_*$ can be taken to be as low as TeV for 
$n\geq 2$. In order to
accommodate the success of the SM and the non-observation of a tower of  
Kaluza-Klein excitations
of SM particles, one can simply confine all fields charged under the 
SM gauge group to a 3D hypersurface.
This can be achieved by D-branes \cite {Pol} of string theory.

A second scenario pioneered by Randall and Sundrum (RS) \cite{RS1} 
solves the hierarchy problem by 
introducing a warp factor to the 4D metric. 
This construction makes use of one extra dimension which is parametrized
by the coordinate $z = r_c\phi$ and the points $(x^{\mu},\phi)$ and $(x^{\mu},-\phi)$ are identified. Here, 
$x^{\mu}$ denotes the usual 4D Minkowski space coordinates. The metric is given by
\beq
\label {RSmetric}
 ds^2 = e^ {-2kr_c |\phi|} \eta_{\mu \nu} dx^{\mu} dx^{\nu} -r_c^2 d\phi 
\eeq
where the metric $\eta_{\mu \nu}$ has the signature (+ - - -). 
The full five dimensional metric will be later referred to as
$G_{AB}$.
The parameter $k$ is a measure of the
curvature of the compactified dimension. For consistency one would expect 
$k \lesssim M_5$ where $M_5$ is
the fundamental 5D scale. The construction also uses two branes, one located at $\phi=0$ 
which is referred to as the hidden
brane and another at $\phi=\pi$ where the 
SM particles are localized. We shall
refer to the latter as the visible brane. The 
effective Planck scale $M_{Pl}$ as seen by observers 
confined to the visible brane is 
\beq
\label{RSPl}
 M_{Pl}^2 = \frac{M_5^3}{k}(1-e^ {-2kr_c\pi}).
\eeq
Unlike the first scenario the radius $r_c$ is not large and the 
hierarchy between the weak and the
Planck scale is solved by taking $kr_c \approx 12$.

While the use of extra dimensions to solve the hierarchy problem is 
very intriguing it also brings
with it many new challenges.  In particular, in the area of neutrino physics 
it abjures the generation of
small neutrino masses via the seesaw mechanism. 
This is because large mass scales are absent from
 the theory and the only operating scale in the visible brane is the 
TeV mass scale.  Instead, in both scenarios neutrino masses are generated   
by allowing the SM singlet fermion, which is usually referred to as
the right-handed neutrino, to be a bulk field just like the graviton. 
Bulk neutrino are discussed in the context of the factorizable geometry
by Arkani-Hamed, Dimopoulos, and Dvali
(ADD) \cite {fgnu}. The smallness of the neutrino mass is due to the scaling of
the Yukawa coupling by the volume of the large extra dimensions. 
Solutions to the atmospheric
neutrino and solar neutrino anomalies are discussed in \cite {DS, ADDos}. 
Further phenomenological studies
can be found in \cite {nufg}.

The active neutrino receives a small mass in a different way in the RS model. The
bulk neutrino has a zero mode with a very small wave function at the visible brane 
\cite {GN}. After 
spontaneous symmetry breaking, with both the Higgs field and the lepton doublet 
residing on the visible
brane, a light neutrino of mass in the range of $\mathrm 10^{-2}$ to 1 eV range 
is generated 
(for details see \cite {GN}). This can be generalized to Higgs fields in the 
bulk \cite {CN}. All of these models for producing neutrino mass in
an extra dimensional scenario represent the simplest constructions and
contain no new additional 
particles and have no new gauge interactions added.
Models in which some or all of the SM fields are allowed to propagate into the 
bulk have also been 
constructed \cite {CM}. It is found that stringent constraints are put upon  
the KK excitations 
of these models from current experiments. However, these limits are 
highly model dependent and it is
difficult to draw general conclusions on the viability of the 
extra dimensional scenarios.

As discussed previously the brane world scenario has a natural setting in 
string theory; however, it 
can be studied from the effective field theory point of view. One
difficulty of such an approach is that we are now dealing with field theories in higher dimensions which
are known to have divergences when virtual loop effects are considered. Such radiative corrections are
well tested in many precision measurements which have firmly established local quantum field theory as 
the correct framework for physics up to the weak scale. Most notable of these is the 
anomalous magnetic 
moments of the muon, $a_{\mu}$, and that of the electron, $a_e$. Currently the world average value for
$a_{\mu}$ is $11659210(46) \times 10^{-10}$ and the difference between the theoretical and experimental value
for $a_{\mu}$ is $\Delta a_{\mu} \equiv a^{exp}_{\mu} - a^{SM}_{\mu} = (43 \pm 45) \times 10^{-10}$.
The E821 experiment at BNL plans to reduce the error to 0.35 ppm \cite {E821}. 
This is clearly an important quantity for constraining any model of physics 
beyond the SM ; and is particularly so
for the effective field theory approach to extra dimensions and 
brane world scenarios.

In this paper we study $a_l$ where $l = \mu, e$ in the both the 
ADD and the RS models 
with bulk neutrinos. We shall ignore the effects of neutrino mixings 
among the three families
and concentrate only on the effects coming from extra dimensions and bulk neutrinos. 
A previous study \cite {Gra}
of $a_{\mu}$ in the ADD model used  additional Higgs doublets and found that the 
dominant effect of bulk neutrinos 
also involved charged Higgs exchange. For the  ADD case  we shall use the minimal 
model where 
 all the SM particles are localized on the brane and only a right-handed neutrino 
is allowed to propagate in the bulk \cite {DS}. We
shall ignore the graviton and its KK excitations as this is discussed in 
\cite {Gra}. For the RS scenario the two models
with bulk neutrinos we use are given in \cite {GN} (brane Higgs) and 
\cite {CN} (bulk Higgs).

It is illustrative to begin our discussions with the gauge invariant 
contribution to $a_l$ from a
 massless neutrino in the SM. The Feynman diagrams are depicted in 
Fig. 1. and the mass insertion  
technique is employed for fermions. A simple calculation 
gives the result in standard notations: 
\beq
\label{ammnu}
 a_l^{\nu}=\frac{G_Fm_l^2}{4\pi ^2 \sqrt{2}}(\frac{5}{3})
\eeq
where the superscript denotes the contribution from the active $\nu_l$ of the SM. 

To see the effects of bulk neutrinos and extra dimensions on the 
anomalous magnetic moment
we begin with  the ADD case, 
with one extra dimension compactified into a circle of radius R. 
Thus the geometry is $M^4 \times T^1$. The
brane where the SM particles reside is located at $z=0$,
i.e. the origin of the extra dimension. The
effective Lagrangian density in 4D involving the active neutrino and the bulk 
neutrinos, $N$, is given by
\beq
\label{FGL}
   {\cal {L}} =\int _0^{2\pi R} dz \bar {N}(i\gamma^{\mu}\partial_{\mu} + i \Gamma_5 \partial_z)N
 + y_* \int_0^{2\pi R} dz \delta(z) \bar{L}H N_R + h.c.
\eeq
where $L$ is the left-handed SM lepton doublet, $H$ denotes the SM Higgs doublet. Our choice of the
Clifford algebra representation, $\Gamma^A$, for 5D spacetime is : $\Gamma^{\mu} = \gamma^{\mu}$ 
for $\mu = 0,1,2,3 $ and $\Gamma ^5 =i\gamma ^5$. The 5D Yukawa coupling, $y_*$, is dimensionful and is related 
to the dimensionless coupling $y$ via
\beq
\label{y5}
    y_*=\frac{y}{M_*^{n/2}}
\eeq
and $n=1$ for 5D. We have neglected a possible higher dimensional bare Dirac mass term for simplicity.
 We implement the Kaluza-Klein ansatz by Fourier expanding 
the fields $N_{L,R}$ as follows:
\beqa
\label{ft}
 N_R &=& \frac{1}{\sqrt{2 \pi R}}\sum_{k = -\infty}^{\infty}n_{kR} e^{\frac {ikz}{R}}, \\
 N_L &=&\frac{1}{\sqrt{2 \pi R}}\sum_{k =-\infty}^{\infty}n_{kL} e^{\frac {ikz}{R}} .
\eeqa
The mass terms necessary for our calculations are obtained by substituting the above into Eq.(\ref{FGL}) and after
electroweak breaking we have
\beq
\label{4Dmass}
 m_D\bar {\nu}_l n_{0R} +m_D\sum_{k=1}^{\infty} \bar{\nu}_{lL}(n_{kR}+ n_{-kR}) +\sum_{k=1}^{\infty} m_k(
\bar{n}_{kL}n_{kR} - \bar{n}_{-kL}n_{-kR}) + h.c.
\eeq
where 
\beq
\label{tw}
m_k=\frac {k}{R} , \,\,\, m_D = \frac {yv}{\sqrt {4\pi R M_*}}
\eeq
and $v=247$ GeV. The 
contribution of the KK tower of bulk neutrino states to $a_l$ which is
first order in $m_D$, comes from calculating
the Feynman diagrams of Fig.2 where the crosses denote mass insertions 
via Eq.(\ref{4Dmass}). It is given by
\beq
\label{abn}
   a_l^{BN}= - \frac{g^2|m_D|^2m_l^2}{16\pi^2}\sum_k \frac{1}{(w^2 -m_k^2)^2}\left [\frac{2}{3}-
\frac{m_k^2(w^2+5m_k^2)}{6w^2(w^2-m_k^2)} -\frac{m_k^2(w^2-2m_k^2)}{(w^2-m_k^2)^2} \ln{\frac{w^2}{m_k^2}} \right ] 
\eeq
where $g$ is the SU(2) gauge 
coupling and $w$ is the mass of the W-boson. As seen from Eq.(\ref{tw}) the KK
states are separated by an equal amount of $1/R$ and this spacing is small if the 
compactification radius 
is relatively large, such as micron size. The infinite sum in Eq.(\ref{abn}) 
can be approximated by an integral 
over the $m_k$. This integral is divergent and has to be cut off at the high 
mass region. A natural choice 
for this cutoff is $M_*$.  When the leading term of the 
integrals is evaluated, the $M_*$ dependence
is compensated  by the corresponding scaling factor in $m_D$ such that the final 
result is finite in the
limit $M_* \rightarrow \infty$. Explicitly, we have 
\beqa
\label{abnall}
 a_l^{BN} & \cong
&
- \frac{5 m_l^2 |y|^2}{3(n-2)2^{2n+3} \pi^{\frac{n+4}{2}}\Gamma (\frac{n}{2})M_*^2} \;\;\;\;\;\;\;  n\neq 2 ,\\
      &\cong&  - \frac{5m_l^2 |y|^2}{768 \pi^3 M_*^2}\ln {\frac{M_*^2}{w^2}}\;\;\;\;\;\;\;\;\;\;\;\;\;\;\;\;\;\;\;\;\;\;\;\;  n=2.
\eeqa
This contribution is negative and subtracts from the SM value except for $n=1$ where it adds.
In these models the effect of the right-handed bulk neutrinos enters
through mass insertions.  An even number of such insertions are required due to the chiral nature
of the SM interactions on the visible brane. This accounts for the factor of $|y|^2$.  
Higher order insertions will be subleading since $m_D$ 
is in general a small quantity. The overall negative sign in Eq.(\ref {abnall}) is a 
result of the loop integration which led to Eq.(\ref{abn}).
We show the result
graphically in Fig. \ref{fig3}, along with the experimental uncertainty on
$a_{\mu}$.  The Yukawa coupling is at $y=1$ as an example. 
The contribution from two extra dimensions is nearly as large
as for one, due to the logarithmic dependence in Eq.(\ref{abnall}).  However,
the contribution is still well below what is measurable currently and in the
near future.  

The anomalous moment of the electron is smaller due to the two powers
of lepton mass in Eq. (\ref{abnall}).  
The contribution to $a_e$ from the bulk
neutrinos is at largest around $10^{-15}$, whereas the experimental
error is at around $10^{-11}$.

The physics of bulk neutrinos in the RS model is different from that of the ADD case. 
We study both of
these cases to examine whether the contribution can be detected in a 
precision measurement of $a_l$.  In the context of the RS scenario,
only 5D models have been studied in any detail.  Below 
we outline the essential steps for obtaining the couplings of bulk 
neutrinos with the brane fermions
which are necessary for our calculations.

The action for the bulk neutrino in the RS model [see Eq.({\ref{RSmetric})] is 
given by \cite {GN}
\beq
\label{RSN}
 S = \int d^4x \int d\phi \sqrt{G} \left [  E_a^A\frac{i}{2} \bar{\Psi} \gamma^a (\overrightarrow {\partial}
 _A - \overleftarrow {\partial}_A)\Psi -m {\rm sgn}(\phi)\bar{\Psi} \Psi \right ] ,
\eeq
where $\gamma^a = (\gamma ^{\mu} , i\gamma ^5)$, $G=\mathrm{det}(G_{AB})$, 
$E_a^A = {\rm diag} (e^{\sigma} , 
e^{\sigma}, e^{\sigma}, 
e^{\sigma}, \frac{1}{r_c})$ is the inverse vielbein, 
$\sigma =kr_c |\phi|$, and $m$ is a Dirac mass. 
We have neglected the 
spin connection term which plays no role in our investigation. The bulk neutrinos are 
Dirac fermions and
the left and right-handed projections come from $\Psi_{L,R} \equiv \frac{1}{2} (1 \mp \gamma^5)$
with the periodic
 boundary condition $\Psi_{L,R}(x, \pi)= \Psi_{L,R}(x, -\pi)$. The KK
decomposition of the $\Psi$ is given by
\beq
\label{KKpsi}
 \Psi_{L,R}(x,\phi) = \sum _n \frac{e^{2\sigma}}{\sqrt{r_c}}\psi_{L,R}^n(x)\hat{f}_{L,R}^n(\phi)
\eeq
and gives rise to  the set of 4D Dirac equations
\beq
\label{bndirac}
 S = \sum_n \int d^4x \left( \bar{\psi}_{L,R} ^n i\gamma^{\mu}\partial_{\mu} \psi_{L,R}^n - m_n 
\bar{\psi}_{R,L}^n  \psi_{L,R}^n  + h.c. \right)
\eeq
with the conditions 
\beq
\label{fncond}
 \int^{\pi}_{0} d\phi e^{\sigma}\hat{f}^{m*}_L\hat{f}^n_L = \int^{\pi}_{0} d\phi e^{\sigma}\hat{f}^{m*}_R \hat{f}^n_R 
= \delta^{mn}.
\eeq
and
\begin{equation} 
\left( \pm {1 \over r_c} \partial_\phi - m \right) 
\hat{f}^{n}_{L,R} + m_n e^{\sigma} \hat{f}^n_{R, L} = 0.
\end{equation} 
For convenience we define the variables 
$\epsilon \equiv e^{-kr_c\pi}, x_n \equiv m_n/(k\epsilon)$,
$t \equiv \epsilon e^{\sigma}$ and the rescaled 
function $\hat{f}^{L,R}_n(\phi) \equiv\sqrt {kr_c \epsilon} f^{R,L}_n(t)$.
We are interested in the mass of the KK modes, $m_n$.  These can be
found by way of the solutions for $f_R$. 
The eigenvalues $m_n$ are determined by the roots of Bessel functions of
order $\nu \equiv m/k$ given below : 
\beq
\label{mn}
      J_{\nu-\frac{1}{2}}(x_n ) = 0.
\eeq
We also note that the value of $f_n^R(1) = \sqrt{2}$ for all $n\neq 0$
at the visible brane can
be obtained from the above equations and the appropriate boundary conditions
which satisfy the orbifold symmetry.

In order to see the implications of Eq. (\ref{mn}) we must know $\nu$ and also
$k$ and $r_c$.  If the bulk neutrino partners with 
the active $\nu_L$ to form a light neutrino then $\nu \simeq 1.1- 1.5$ \cite {CN}. 
The KK bulk neutrino states
are approximately equally spaced as the roots of Eq.(\ref{mn}).  They are given 
approximately by $n\pi$ for $\nu \approx 1$.
Since the solution of the hierarchy problem requires 
$kr_c \approx 12$ and also $k \lesssim  M_5$, then the K.K. states will be of 
order the weak scale and the bulk Dirac mass $m \sim M_5$.  This is independent of
whether the Higgs doublet resides only on the visible brane or in the bulk.

To proceed we introduce the Higgs doublet which couples the bulk field 
to the chiral lepton
doublet which is confined on the brane. We first examine the case which has 
this scalar field, 
$h_0(x)$, also localized on the visible brane. The relevant action terms are 
\beq
\label{4Dhiggs}
  S = \int d^4x \epsilon^4 \left [ e^{2kr_c\pi}\partial ^{\mu}h^{\dagger}_0 \partial_{\mu}h_0 -
\lambda (|h_0|^2 -\frac{v_0^2}{2})^2  - (\frac{Y_5}{\sqrt{M_5}}\bar{L}_0(x)h_0(x)\Psi_R(x, \pi)+h.c.) \right ]  ,
\eeq
where the subscript 0 denotes bare fields and ${Y}_5$ is the dimensionless  
Yukawa coupling. As noted in \cite {RS1},
in order to get the canonical normalization of fields on the brane one must 
do the rescalings:
$h_0 \rightarrow h e^{kr_c\pi}$ and $ L_0 \rightarrow e^{\frac{3}{2}kr_c\pi}L$ 
and obtain $v=v_0\epsilon$
which we identify as the weak scale.  After electroweak symmetry breaking, 
 we can use Eqs. (\ref{KKpsi}, \ref{4Dhiggs})
 and the rescaling relations, to find  
the coupling between $\nu_{lL}$ and the 
$n^{\mathrm {th}}$ KK bulk neutrino:
\beqa
\label{GNin}
   y_n v &=& Y_5v\sqrt {\frac{k}{2M_5}}f^R_n(1)   \\
         &=& Y_5v \sqrt {\frac{k}{M_5}}
\eeqa
for $n\neq 0$. The zero mode coupling is very small and is 
interpreted as the light neutrino mass. On the other hand, 
the mass of the KK excitations are of the weak scale. 
Taking $\nu \simeq 1.1$ and $k \sim 10^{18}$ GeV as an
example, we obtain $m_1 \sim 133$ GeV and $m_2 \sim 266$ GeV etc. 
Thus we can expect most KK modes  to be heavier
than the W boson. The contribution to $a_l$ from the bulk neutrinos in the RS model with brane Higgs can now
be obtained from Eq.(\ref{abn}). Keeping only the leading term we find
\beqa
\label{arsh}
 a_l^{RS}&\simeq& \frac{|Y_5|^2 m_l^2k}{4\pi^2 M_5}(-\frac{5}{6})\sum_{n=1}^{\infty}\frac{1}{m_n^2}  \\
          &\simeq& -\frac{5|Y_5|^2m_l^2}{24\pi^4 k M_5\epsilon^2}\sum _{n=1}^{\infty} \frac{1}{n^2} \;\;\;\;\;\;\;\;
\;\;\;\;\;( \mathrm {for} \;\nu=1.1) \nonumber \\
          &=& -\frac{5|Y_5|^2m_l^2}{144 \pi^2 k M_5\epsilon^2} ,
\eeqa
where we have taken the sum to infinitely many KK states.  The masses of the
KK modes are given by the roots of the Bessel function.  We approximate 
the roots as being spaced by $\pi n$.  This approximation is best for the
lower values of $\nu$.  For example, for $\nu \approx 1.5$ they
are spaced as $(n + 5/4) \pi$, so for this case the result is slightly
smaller than in the expression above.  Figure \ref{fig4} shows the magnitude 
of the correction to $a_\mu$ from the bulk neutrinos in the RS model.
For the illustrative purposes of this figure, we have taken $k \epsilon \sim v$ and also
$M_5 \epsilon \sim v$.  Again, the contribution to $a_e$ is several orders
of magnitude below current experimental uncertainty due to the
$m_e^2 / (k M_5 \epsilon^2)$ suppression.

It was pointed out in \cite {CN} that the phenomenology of the 
SM on the visible brane depends on
whether the Higgs field is allowed to propagate in the bulk or not. 
Here we explore the consequences of
bulk Higgs on $a_l$. The action of the bare bulk Higgs field, $H_0$ 
we are interested in is given by
\beqa
\label{BHA}
S_H &=& \int d^4x \int_{-\pi}^{\pi} d\phi \sqrt{G} \left [G^{AB}D_A H_0^{\dagger}D_B H_0 -\frac{\lambda_B}{4M_5}
\left( H_0^{\dagger}H_0 -\frac{v_0^3}{2}\right)^2 \right ]    \nonumber \\
     &&-\left[ {Y_B \over  M_5} \int d^4x \epsilon^4 \bar{L}_0 H_0(x,\pi)
\Psi_R(x,\pi) +h.c. \right]
\eeqa
where the Yukawa coupling $Y_B$ is dimensionless, and $D_A$ is the gauge covariant 
derivative. Spontaneous
symmetry breaking is achieved by $H_0 \rightarrow (H+v_0^{\frac{3}{2}})/\sqrt{2}$. 
The weak scale is obtained
from the gauge terms (see \cite {CN}) and is
\beq
\label{fermi}
  v=\epsilon v_0\sqrt{r_cv_0}
\eeq
The mass insertion can be obtained from the Yukawa term of Eq.(\ref{BHA}), Eq.(\ref{KKpsi}), and the field rescaling.
Explicitly the $\nu_{L}\psi_n$ coupling is given by
\beq
\label {bhnp}
  \frac{1}{\sqrt{2}}Y_B v_0 \sqrt{v_0 k} { \epsilon \over M_5}  f_n^R(1)= Y_B v 
\sqrt{k \over M_5} \left( {1 \over \sqrt{ M_5 r_c}} \right)
\eeq
for all KK bulk neutrinos.  This has a suppression factor of $ 1 / \sqrt{ M_5 r_c}$,
as compared with the brane Higgs case.  This suppression factor occurs
in all bulk Higgs Yukawa couplings, and is not 
unique to the coupling of the bulk neutrinos.  For example it also
appears in brane fermion mass terms, such as the electron mass.  
This factor is interpreted
as a general rescaling of the Yukawa coupling.  Therefore the
expression for $a_l$ is the same as in the brane Higgs case except
for the rescaling of the coupling.

We have examined the effect of virtual bulk neutrinos by calculating the
correction to the anomalous magnetic moment of the charged leptons.  We
have employed three different models: a factorizable geometry with up to
six extra dimensions (ADD), and two versions of the Randall-Sundrum scenario
with one extra dimension.  In each case we take the simplest model, with
only the bulk neutrino and/or the Higgs field propagating in the bulk.
This demonstrates the key features of the extra dimensional neutrino
scenarios.  A Kaluza-Klein tower of bulk neutrinos must be summed
 in order to calculate the correction. In all cases
 the correction is finite and does not diverge as the
higher dimensional scale increases, which corresponds to taking the KK tower to
be infinite. This indicates that the predictions are robust.
It is interesting to note how the new physics is probed by comparing
the Eqs. (\ref{ammnu}, \ref{abnall}), and (\ref{arsh}). In the SM the size of $a_l$ 
is set by the weak scale whereas in the ADD case this is replaced by the scale $M_*$.
On the other hand in the RS model the determining scale is the redshifted scale $M_5\epsilon$. 
In addition, the contribution to the electron magnetic moment
is significantly smaller than
that of the muon, because of suppression factor, $(m_l / M_*)^2$ or 
$(m_l / \epsilon k)^2$.  In all scenarios, the contribution from the 
bulk neutrino modes
is currently below experimental precision. However the Randall-Sundrum
models are close to being probed by the next generation
of g-2 measurements for the muon.

        While this paper was being written a preprint appeared that investigated $a_l$ in the
RS scenario with the SM chiral fermions and gauge bosons all allowed to propagate in the bulk \cite {dhr2}.   

        This work is partially supported by the Natural Science and Engineering Research Council
of Canada

\newpage


\begin{figure}
\epsfxsize=14cm
\epsfxsize=14cm
\caption{Standard Model contribution to g-2 of the electron in the
unitary gauge, which has a 
neutrino as an intermediate state.  The cross represents a mass insertion, which
can be on either external line.}
\label{fig1}
\end{figure}

\begin{figure}
\epsfxsize=14cm
\epsfxsize=14cm
\vskip 0.5in
\caption{ First order bulk neutrino 
contribution to g-2 of the electron in the unitary gauge.
As in Fig. \ref{fig1} there is an additional diagram with the electron mass 
insertion on the other external line.
}
\label{fig2}
\end{figure}

\begin{figure}
\epsfxsize=14cm
\epsfxsize=14cm
\vskip 0.5in
\caption{Shows the predicted contribution to g-2 for the muon 
from the bulk neutrinos for
different numbers of extra dimensions in the ADD scenario.   The horizontal 
dashed lines show the current (upper) and future expected (lower) experimental
sensitivities.   The solid line shows the contribution for one extra dimension, while
the dot-dashed line shows the contribution for two extra dimensions.  The 
lowest dashed line on the plots shows the contribution for three extra dimensions.}
\label{fig3}
\end{figure}

\begin{figure}
\epsfxsize=14cm
\epsfxsize=14cm
\vskip 0.5in
\caption{Shows the predicted contribution to g-2 from the bulk neutrinos for
the Randall-Sundrum scenario.  The dashed horizontal lines show experimental
sensitivity as in Fig \ref{fig3}.  The solid line shows the contribution 
from bulk neutrinos when the Higgs boson is restricted to the 3+1 dimensional brane.  
}
\label{fig4}
\end{figure}

\newpage

\begin{figure}
\epsfxsize=14cm
\epsfxsize=14cm
\centerline{\epsfig{file=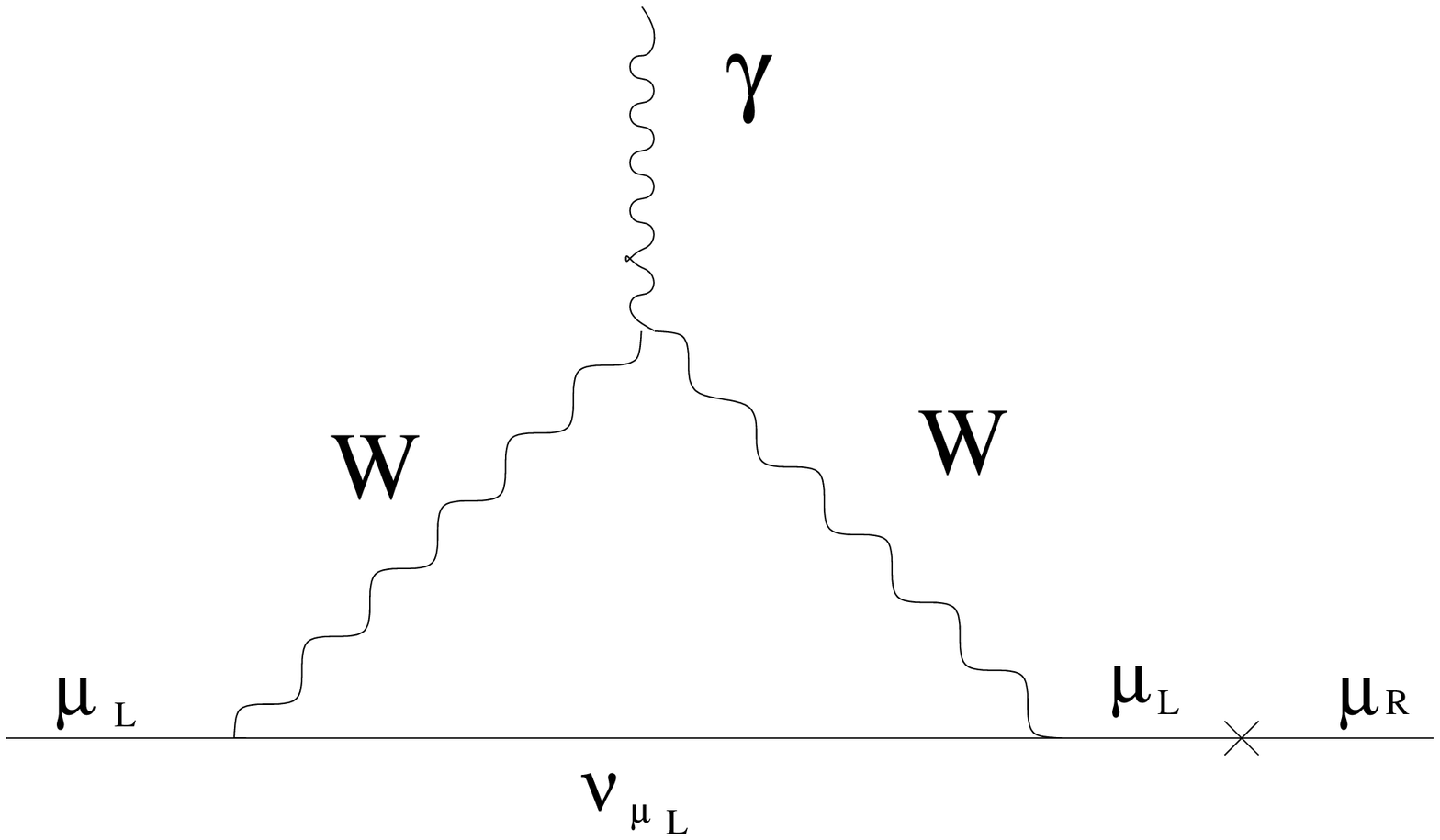,angle=-90}}
\end{figure}

\begin{figure}
\epsfxsize=14cm
\epsfxsize=14cm
\centerline{\epsfig{file=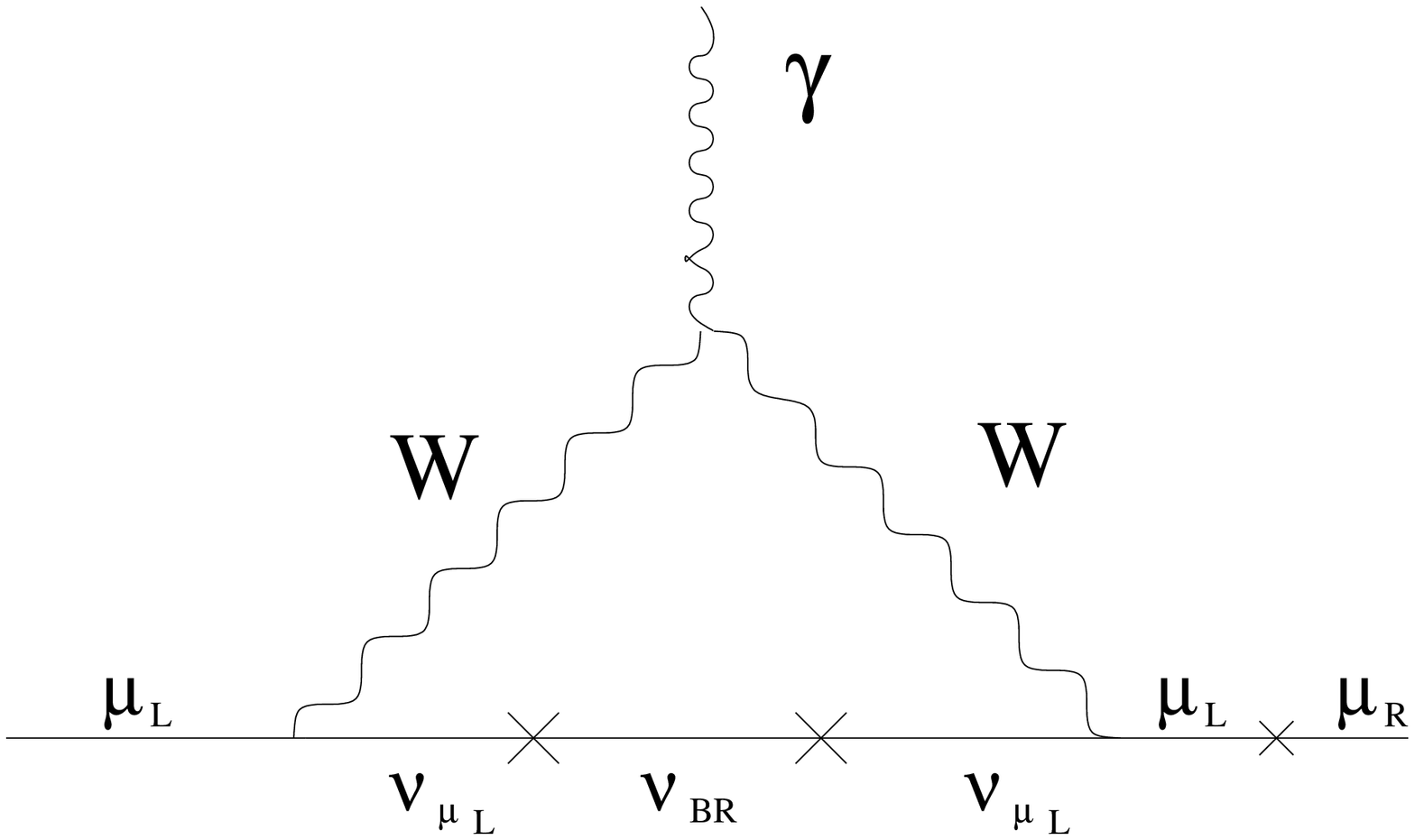,angle=-90}}
\end{figure}

\begin{figure}
\epsfxsize=14cm
\epsfxsize=14cm
\centerline{\epsffile{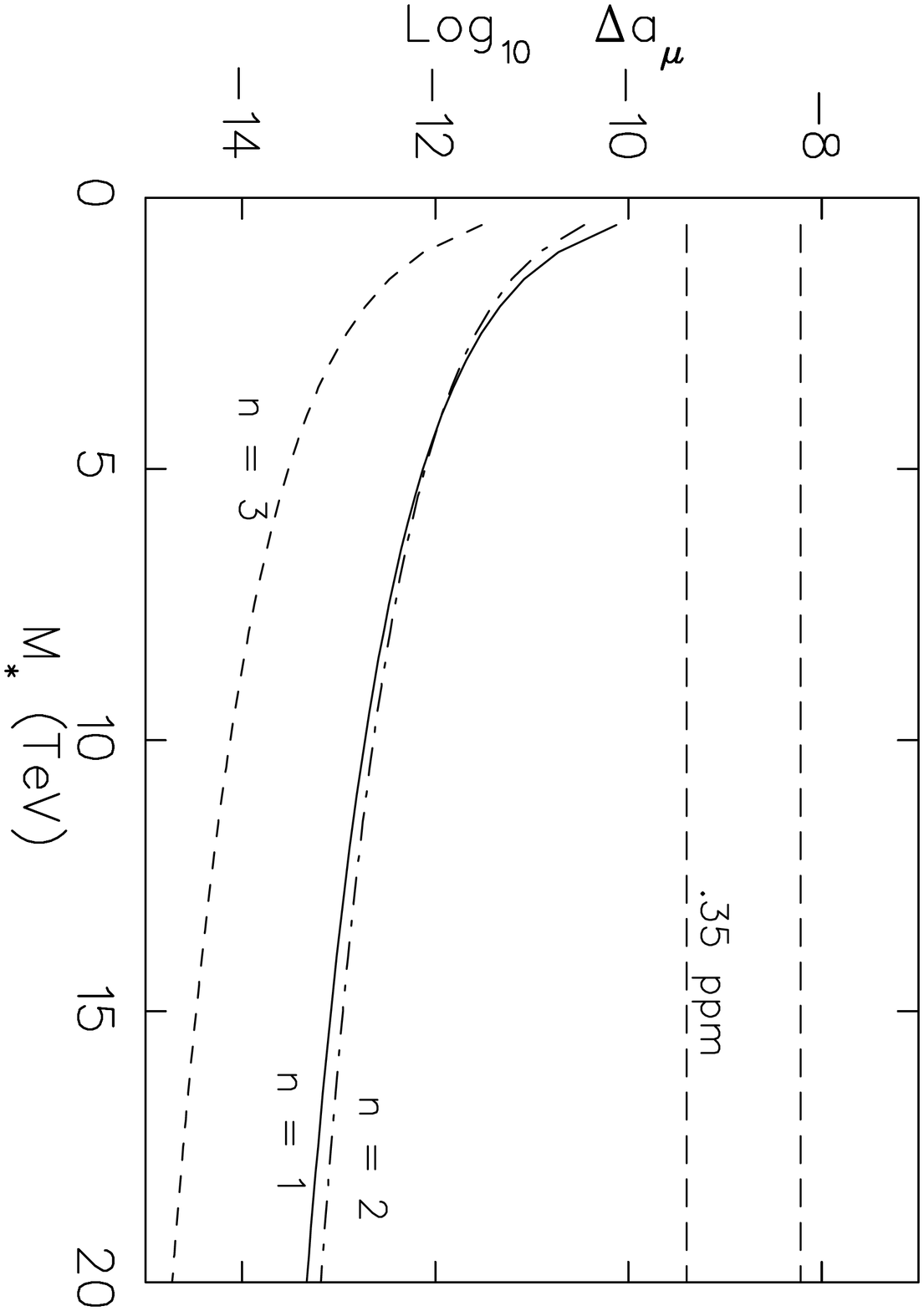}}
\end{figure}

\begin{figure}
\epsfxsize=14cm
\epsfxsize=14cm
\centerline{\epsffile{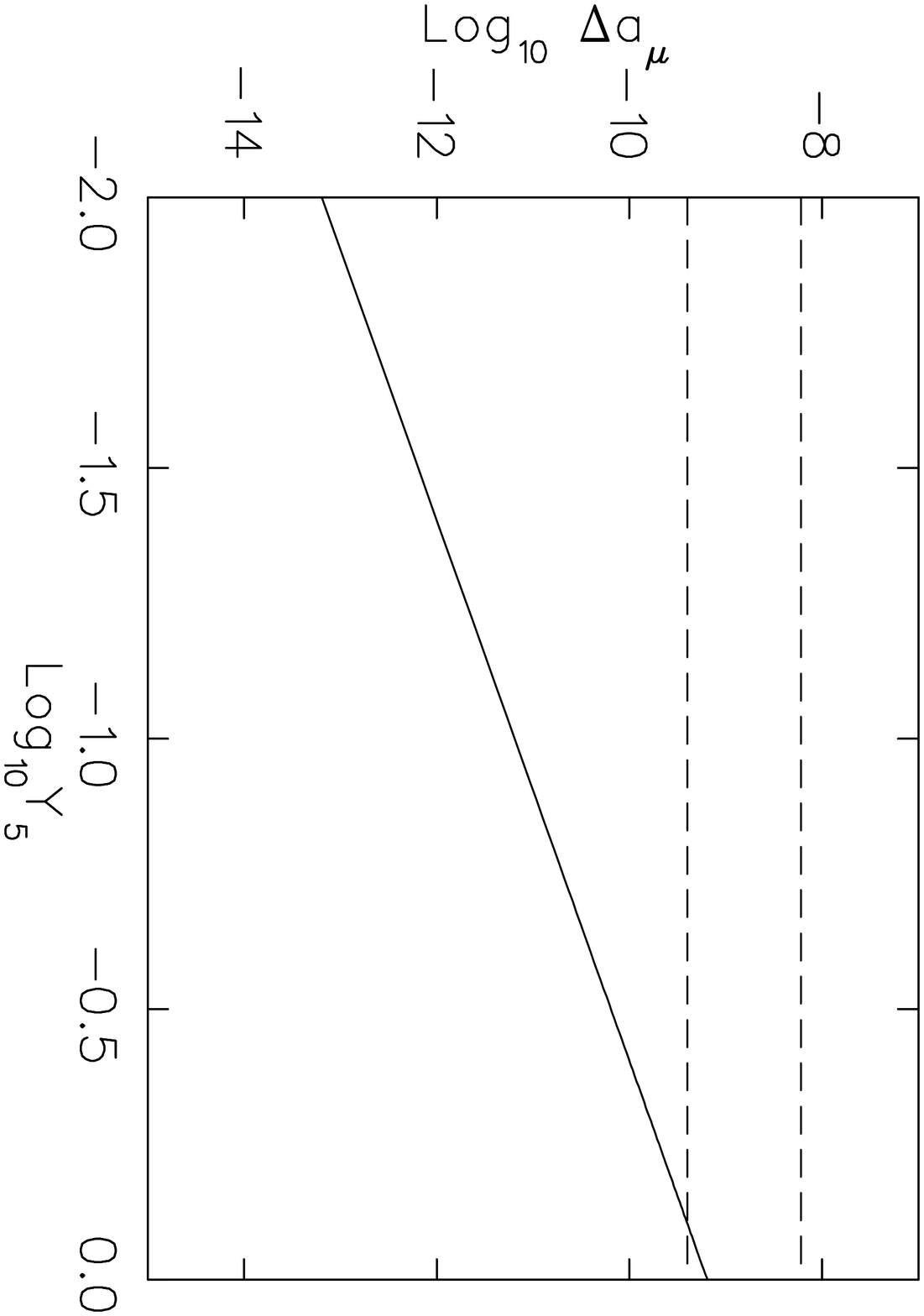}}
\end{figure}

\end{document}